\documentclass[journal,10pt]{IEEEtran}

\usepackage{amsfonts}
\usepackage{amsmath}
\usepackage{graphicx}
\usepackage{booktabs}
\usepackage{url}
\usepackage{xcolor}
\makeatletter
\let\MYcaption\@makecaption
\makeatother

\usepackage[font=footnotesize]{subcaption}

\makeatletter
\let\@makecaption\MYcaption
\makeatother

\begin{document}

\title{Implicit Neural Representation for 3D Radio Map Compression in Low-Altitude Environments}

\author{Jiaming Cheng,~\IEEEmembership{Student Member,~IEEE,}
        Wei Chen,~\IEEEmembership{Senior Member,~IEEE,}
        Yuchen Li,\\
        Bo Ai,~\IEEEmembership{Fellow,~IEEE}
        and Jun Zhang,~\IEEEmembership{Fellow,~IEEE}
\thanks{Jiaming Cheng, Wei Chen, Yuchen Li, and Bo Ai are with the School of Electronic
and Information Engineering, Beijing Jiaotong University, China (e-mail:  \{jiamingcheng, weich, 23331187, boai\}@bjtu.edu.cn). }
\thanks{Jun Zhang is with the Department of Electronic and Computer Engineering, Hong Kong University of Science and Technology, Hong Kong (e-mail: eejzhang@ust.hk).}
\thanks{The corresponding author is Wei Chen.}}
%\textit{(Corresponding author: Wei Chen.)}
\maketitle

\begin{abstract}
Low-altitude wireless networks require accurate 3D radio maps (RMs) to support environment-aware communication. However, due to the extensive spatial coverage and multi-modal channel representations, 3D RMs usually incur prohibitive storage and transmission overhead, making efficient compression highly desirable. In this paper, we propose NeRM3D, an implicit neural representation (INR)-based framework for compact 3D RM compression. Specifically, NeRM3D represents the entire 3D RM as a lightweight neural network that maps discrete height values to corresponding RM slices, thereby encoding the RM into a compact neural network. To effectively model the volumetric spatial structure of 3D RMs, NeRM3D jointly integrates horizontal embeddings with height-aware modulation, enabling efficient extraction and fusion of spatial correlations across different vertical layers and channel modalities. Experimental results demonstrate that NeRM3D achieves competitive compression performance with low complexity and decoding latency, making it suitable for resource-constrained devices. 
\end{abstract}


\IEEEpeerreviewmaketitle

\section{Introduction}

\IEEEPARstart{T}{he} evolution toward next-generation (NextG) wireless networks is extending terrestrial communications into low-altitude environments~\cite{chen2026toward}. Emerging low-altitude intelligent transportation applications, such as unmanned aerial vehicles (UAVs) and electric vertical take-off and landing (eVTOL) aircraft, impose stringent requirements on high data rates and low-latency communications~\cite{wu2025low}. Conventional geometric 3D environment maps are insufficient for such scenarios, as they cannot provide electromagnetic signal coverage information. In this context, 3D radio maps (RMs), which characterize the spatial distribution of channel features, have emerged as a key enabler for real-time sensing, prediction, and control~\cite{hu20233d}. By explicitly modeling the vertical spatial dimension, 3D RMs will play a critical role in low-altitude wireless networks~\cite{chen2025high}. For example, in UAV-assisted communications and air-ground integrated networks, link quality varies significantly with altitude, requiring accurate 3D channel knowledge for trajectory planning and mobility management~\cite{zeng2024tutorial}.

To date, most existing studies on 3D RMs have focused on map construction, including ray-tracing-based generation~\cite{hoydis2023sionna}, learning-assisted inference from sparse observations~\cite{wang2025radiodiff,chen2026radiogen3d}, and optics-inspired reconstruction methods~\cite{zhao2023nerf2,wen2025neural,wen2026bridging}. Moreover, RadioLAM employs a large generative model to construct fine-grained 2D RMs at a given altitude from sparse measurements and environmental information~\cite{liu2026fine}. These approaches aim to improve the accuracy of 3D RMs under limited sensing resources and have demonstrated promising performance. However, the resulting 3D RMs often exhibit rapidly growing data volumes with increasing coverage area and spatial granularity, leading to substantial storage and transmission overhead. Therefore, it is essential to develop efficient compression techniques to reduce storage costs and enable practical deployment of 3D RMs in real-world systems.

The compression of 3D RMs faces several key challenges. (i) In the use case where a base station (BS) delivers RMs to user equipment (UEs), UEs are typically constrained in computational capability, memory capacity, and energy budget, which necessitates highly lightweight reconstruction models. (ii) In low-altitude scenarios, high mobility imposes stringent requirements on ultra-low decoding latency to support real-time trajectory planning and instantaneous decision-making. (iii) Since 3D RMs characterize the electromagnetic environment of specific geographic regions, they exhibit strong site-specific characteristics, making it difficult for autoencoder (AE)-based compression models to generalize to unseen environments. This challenge is further exacerbated by the scarcity of large-scale, high-fidelity 3D RM datasets, as their acquisition is both costly and time-consuming.

To address these challenges, implicit neural representation (INR) offers a promising solution~\cite{sitzmann2020implicit}. Unlike conventional explicit representations that store discrete pixels, INR represents target signals as continuous functions parameterized by neural networks, thereby reformulating compression as the encoding of network parameters. Owing to their strong representational flexibility, INRs have been successfully applied in image~\cite{dupont2021coin}, video~\cite{chen2021nerv,zhang2024boosting}, and point cloud~\cite{bian2025over} compression tasks. However, existing studies primarily focus on natural image and video representations, while the application to 3D RM compression remains largely underexplored.

In this paper, we develop an INR-based framework for 3D RM compression by exploiting the intrinsic volumetric spatial characteristics of wireless propagation environments. Unlike conventional approaches that treat 3D RMs as collections of independent horizontal slices across different heights, the proposed NeRM3D represents the entire 3D RM using a height-conditioned INR, where all information is implicitly embedded in the network architecture and its parameters. The main contributions of this paper are listed as follows:
\begin{itemize}
\item We propose NeRM3D, an INR-based 3D RM compression framework that reformulates encoding as model fitting and simplifies decoding to lightweight forward inference. Given a training height, NeRM3D reconstructs the corresponding 2D RM slice.

\item We design height-aware conditional convolution blocks, named CondConv, that progressively reconstruct RM slices from an initial low-resolution representation. By jointly integrating height modulation with horizontal embeddings, CondConv blocks effectively capture volumetric spatial correlations across channel modalities.

\item Experimental results demonstrate that NeRM3D achieves competitive rate–distortion performance with low decoding complexity, making it suitable for resource-limited devices in low-altitude wireless networks.
\end{itemize}

\section{System Model}
This section introduces the 3D RM representation and the INR-based compression framework shown in Fig.~\ref{Fig: inr}.
\subsection{3D Radio Map Representation}
A typical 3D RM is represented as a multi-modal tensor, which characterizes the spatial distribution of diverse wireless channel properties. 
In this work, we denote the 3D RM as $\mathcal{R} \in \mathbb{R}^{H \times W \times D \times C}$, where $H$ and \(W\) denote the two planar dimensions, \(D\) denotes the height dimension, and $C$ denotes the number of channel-related modalities. Specifically, each voxel in the 3D grid is associated with a $C$-dimensional vector that describes channel characteristics at that spatial location, such as pathloss, delay, and direction of arrival (DoA) in azimuth and elevation. We assume a perfectly constructed 3D RM $\mathcal{R}$ that faithfully reflects the underlying wireless environment, free from estimation or construction errors.

To ensure consistency across different spatial locations for stable model training, all channel parameters are globally normalized to the range of [0, 1]~\cite{wang2025radiodiff}.  However, storing such dense 3D RMs incurs significant memory overhead, thereby motivating the need for efficient compression techniques.

To characterize vertical spatial dependency in 3D RMs, we evaluate the correlation between horizontal slices at different heights. Let $R_{n,c}(x,y,d)$ denote the $c$-th channel modality at location $(x,y,d)$ in the $n$-th RM, with spatial mean
\begin{equation}
\label{eq:mean}
\mu_{n,c,d}=
\frac{1}{HW}
\sum_{x=1}^{H}
\sum_{y=1}^{W}
R_{n,c}(x,y,d).
\end{equation}

For slices separated by $\ell$ height intervals, the correlation is
\begin{equation} \label{eq:vertical_corr}  \rho_{n,c}(d,d+\ell) = \frac{ \left\langle \widetilde{\mathbf R}_{n,c,d}, \widetilde{\mathbf R}_{n,c,d+\ell} \right\rangle_F }{ \left\| \widetilde{\mathbf R}_{n,c,d} \right\|_F \left\| \widetilde{\mathbf R}_{n,c,d+\ell} \right\|_F }, \end{equation}
where $\widetilde{R}_{n,c}(x,y,d)=R_{n,c}(x,y,d)-\mu_{n,c,d}$ denotes the mean-centered slice, and $\langle\cdot,\cdot\rangle_F$ and $\|\cdot\|_F$ denote the Frobenius inner product and norm, respectively. The correlation is then averaged over all valid height pairs and RM samples for each $\ell$, with $95\%$ confidence intervals computed across RMs.

\begin{figure}[t] 
\centering 
\includegraphics[width=0.36\textwidth]{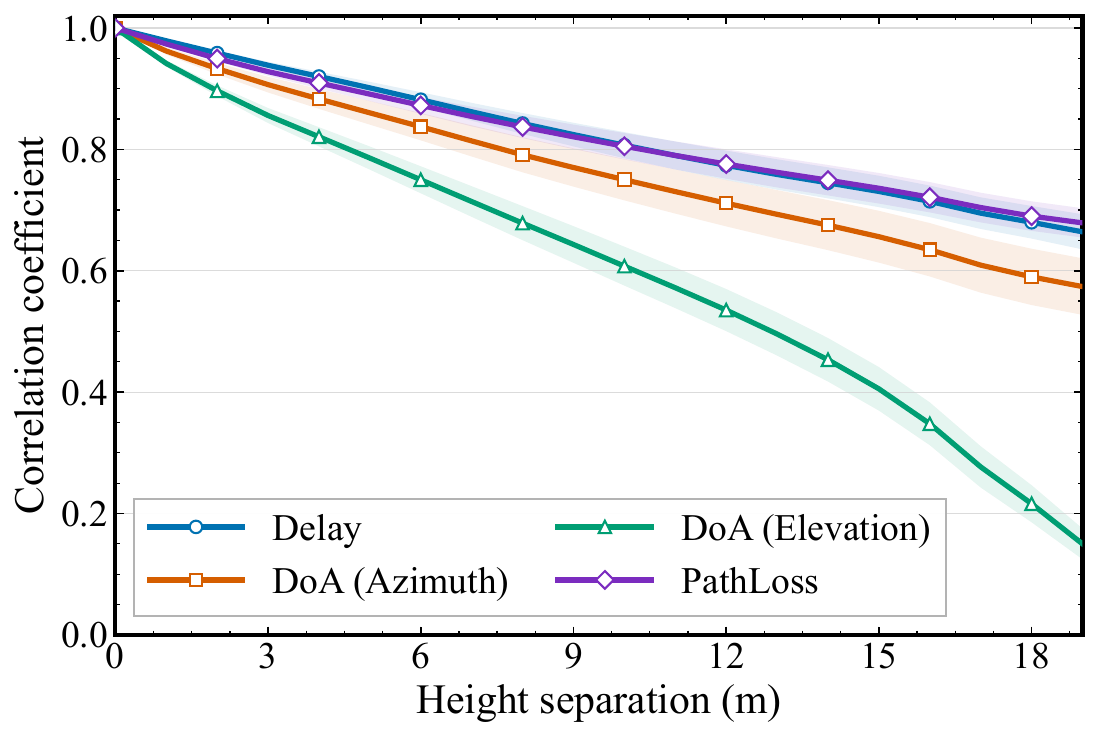} 
\caption{Average vertical correlation versus height separation for different channel modalities.}
\label{fig:vertical_corr} 
\end{figure}

As shown in Fig.~\ref{fig:vertical_corr}, the 3D RMs exhibit strong vertical correlation between nearby height slices, indicating substantial redundancy along the height dimension. The correlation gradually decreases as the height separation increases, with distinct decay patterns across different channel modalities. This vertical dependency motivates the height-conditioned representation and height-wise modulation adopted in NeRM3D.

\subsection{INR-Based Radio Map Compression Framework}

\begin{figure}[!t]
\centering
\includegraphics[width=0.43\textwidth]{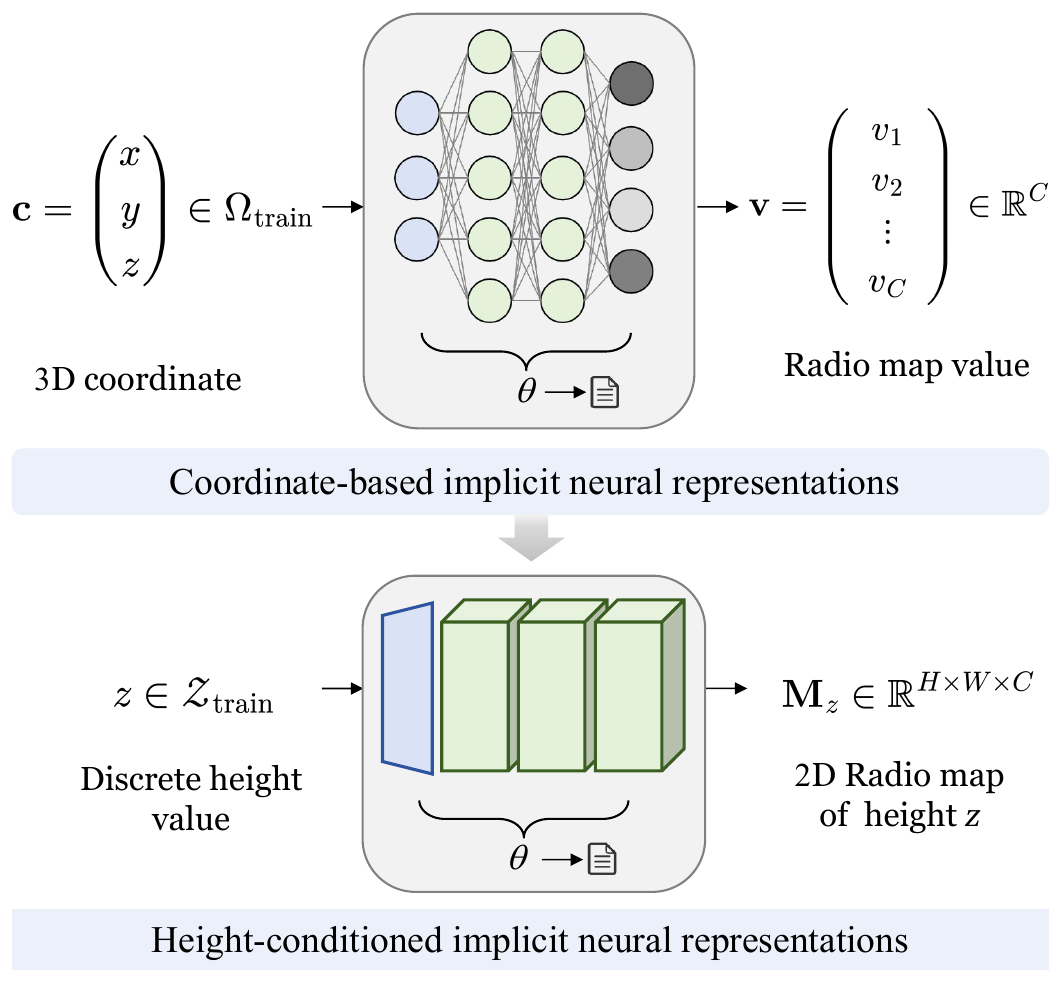}
\caption{Comparison of different INR frameworks for 3D radio map.}
\label{Fig: inr}
\end{figure}
% (a) Coordinate-based INR takes 3D spatial coordinates as input and outputs multi-modal channel properties at each location. (b) The proposed NeRM3D is a height-modulated INR that takes the height index as input and outputs a 2D radio map slice for each height level.

\begin{figure*}[!t]
\centering
\includegraphics[width=0.99\textwidth]{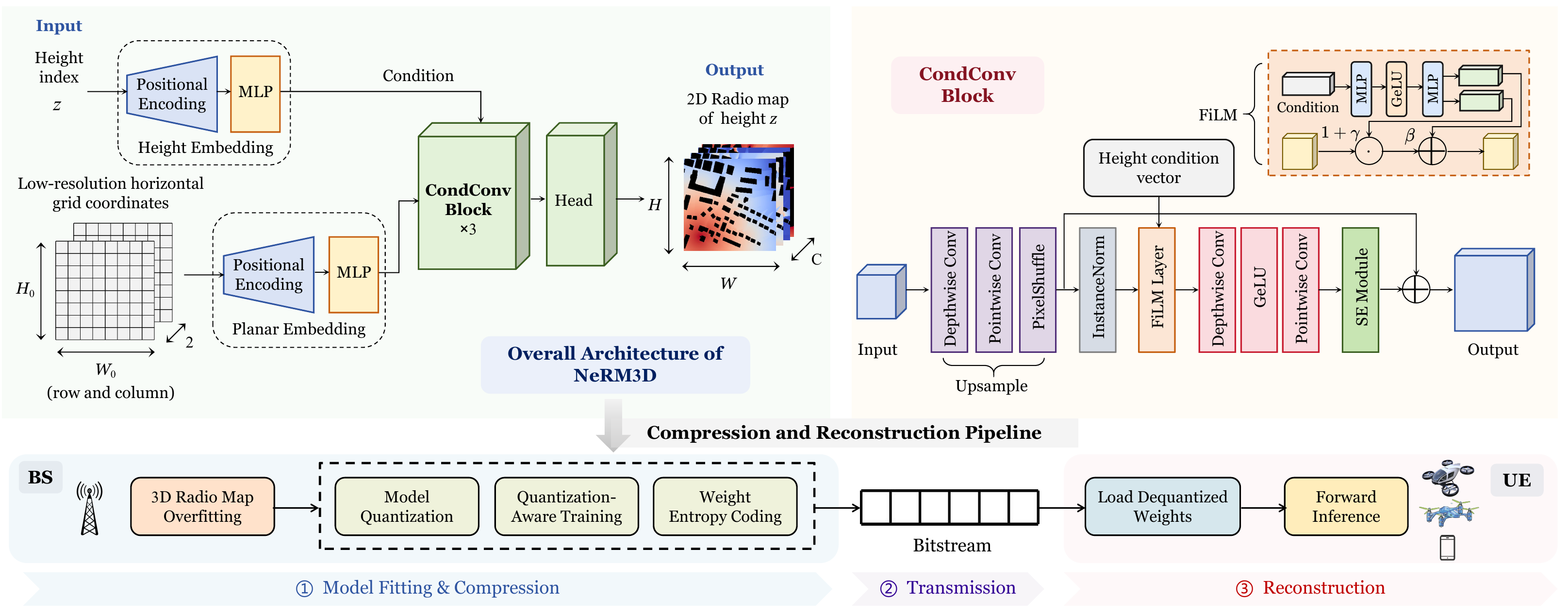}
\caption{Overview of the proposed NeRM3D framework, including the network architecture, CondConv block, and site-specific 3D radio map compression and reconstruction pipeline.}
\label{Fig_NeRM3D}
\end{figure*}

We consider 3D RM compression by reformulating it as a model compression problem. 
Specifically, a lightweight neural network is overfitted to represent the RM, and its quantized weights are transmitted as the compressed bitstream. At the decoder, the RM is reconstructed by querying the network instantiated with the received weights. The compression efficiency is thus determined by the compactness of the model representation rather than the latent dimension.

In this work, we adopt a sinusoidal representation network, SIREN~\cite{sitzmann2020implicit}, as the baseline INR model, as illustrated in Fig.~\ref{Fig: inr}. The model learns a continuous mapping from sampled 3D spatial coordinates $\mathbf{c}$ to the RM values $\mathbf{v}$:
\begin{equation}
f_{\theta}: \mathbf{c}=(x,y,z)\in\Omega_{\mathrm{train}}\subset\mathbb{R}^3 \rightarrow \mathbf{v} \in \mathbb{R}^C,
\end{equation}
where  $\Omega_{\mathrm{train}}$
denotes the predefined 3D spatial sampling grid. Compared with conventional multilayer perceptrons (MLPs), SIREN leverages sinusoidal activation functions, which are more effective in capturing high-frequency spatial variations.

However, such coordinate-based INR requires querying the network at every spatial location, resulting in a large number of forward evaluations for volumetric data. More importantly, it does not explicitly exploit the structural correlations across different height slices, leading to redundant representations. To address these limitations, inspired by neural representations for videos~\cite{chen2021nerv}, we reformulate the 3D RM as a height-conditioned function
that maps each discrete height value
$z$ to its corresponding 2D RM slice $\mathbf{M}_z$,
as illustrated in Fig.~\ref{Fig: inr}:
\begin{equation}
f_{\theta}: z\in\mathcal{Z}_{\mathrm{train}}
\subset\mathbb{R} \rightarrow \mathbf{M}_z
\in \mathbb{R}^{H \times W \times C},
\end{equation}
where $\mathcal{Z}_{\mathrm{train}}$ denotes the set of
sampled height values in $\Omega_{\mathrm{train}}$,
and all slices share the same network
parameters $\theta$.

Based on this formulation, we propose a height-conditioned INR-based network, termed NeRM3D, which integrates horizontal spatial embedding and height-wise embedding to jointly model the 3D structure. The detailed network architecture is presented in Section~\ref{sec_Framework_Design}.
The network is trained using the mean squared error (MSE) loss between reconstructed and ground-truth RM slices:
\begin{equation}
\mathcal{L}=\frac{1}{D}\sum_{d=1}^{D}
\left\|f_{\theta}(z_d)-\mathbf{M}_{z_d}\right\|_2^2.
\end{equation}

Compared to coordinate-based INR that queries all spatial locations, NeRM3D operates on a per-height-slice basis, significantly reducing the number of network queries. This design better exploits spatial redundancy, improving parameter utilization, especially for large-size 3D RMs.

\section{Proposed Network}\label{sec_Framework_Design}
This section presents the proposed NeRM3D architecture, CondConv block, and model compression strategy.

\subsection{Overall Architecture of NeRM3D}
The NeRM3D architecture, illustrated in Fig.~\ref{Fig_NeRM3D}, takes the height index as the network input and reconstructs the corresponding RM slice at the target height. Unlike conventional 3D INR methods, NeRM3D adopts a disentangled formulation for the vertical and horizontal dimensions, reflecting their distinct spatial resolutions and propagation characteristics. The network begins with two parallel encoding branches. Specifically, the height value is mapped to a global height embedding through positional encoding (PE) and an MLP, while a fixed low-resolution planar coordinate grid is encoded as the initial spatial feature map. During forward propagation, the planar embedding is progressively upsampled through a series of CondConv blocks, while the height-conditioned feature vector is used to modulate the feature distribution across decoding stages. Finally, the output head maps the decoded features to the reconstructed RM at the original spatial resolution.

The process begins by separately encoding the height value $z$ and a low-resolution planar coordinate grid $\mathbf{S}_0 \in \mathbb{R}^{H_0 \times W_0 \times 2}$, which is used to initialize the spatial feature context. Specifically, $\mathbf{S}_0$ is constructed by generating a normalized 2D coordinate grid spanning the planar domain of interest, where the horizontal and vertical coordinates are uniformly sampled and normalized to the range $[0,1)$. Taking the RM with a spatial resolution of $256 \times 256$ as an example, both $H_0$ and $W_0$ are set to 32. To capture high-frequency spatial variations, distinct PE configurations are adopted for the height and planar inputs, each with its own frequency bases and frequency levels. Given a normalized scalar input $v\in[0,1]$, the frequency PE is defined as
\begin{equation}
\begin{aligned}
\gamma(v) = [&\sin(\beta^0 \pi v), \cos(\beta^0 \pi v), \dots, \\
            &\sin(\beta^{L-1} \pi v), \cos(\beta^{L-1} \pi v)],
\end{aligned}
\end{equation}
where $\beta$ denotes the frequency base and $L$ is the number of frequency levels. Specifically, different frequency settings are adopted for the height and planar embeddings to capture spatial variations at their respective characteristic scales.

The encoded height value is then processed by an MLP with SINE activation to generate a height condition vector $\mathbf{v}_z \in \mathbb{R}^{d_h}$, while the encoded planar coordinates are projected into a planar feature embedding $\mathbf{S}_{xy} \in \mathbb{R}^{H_0 \times W_0 \times d_h}$. The resulting feature $\mathbf{S}_{xy}$ serves as the initial state and is progressively refined through three CondConv blocks, incorporating modulation from the height condition vector $\mathbf{v}_z$, followed by a reconstruction head that outputs the RM slice.

\subsection{CondConv Block}
To progressively upscale feature maps and reconstruct the RM, CondConv blocks are designed, as illustrated in Fig.~\ref{Fig_NeRM3D}. These blocks incorporate height-conditioned modulation to guide feature transformations across decoding stages. Each block first increases the spatial resolution using a PixelShuffle layer, preceded by a separable convolution for efficient spatial feature extraction with a reduced number of parameters. This is followed by an instance normalization layer to stabilize the feature statistics across different height slices.

To inject height-specific information at each stage of the decoder, inspired by~\cite{perez2018film}, we introduce a feature-wise linear modulation (FiLM) layer to fuse spatial features with height-conditioned embeddings. At decoder stage \(s\), the conditioning MLP maps the height embedding \(\mathbf{v}_z\) to channel-wise modulation parameters \(\boldsymbol{\gamma}_s(\mathbf{v}_z)\) and \(\boldsymbol{\beta}_s(\mathbf{v}_z)\) for feature-wise scaling and shifting, respectively. For feature channel \(c\), the FiLM modulation is given by
\begin{equation}
\widetilde{F}_{s,c}(x,y,z)=
\left[1+\gamma_{s,c}(\mathbf{v}_z)\right]F_{s,c}(x,y)
+\beta_{s,c}(\mathbf{v}_z).
\end{equation}
Consequently, each latent feature channel can learn a different vertical response at every modulated decoder scale.

Following modulation, separable convolution and GELU activation are employed for efficient spatial feature extraction. A squeeze-and-excitation (SE) module further recalibrates channel-wise responses, while a residual connection facilitates stable training and feature propagation.

\subsection{Model Compression}
At the completion of training for a single 3D RM, the model parameters are represented in 32-bit floating-point format. To further reduce storage overhead, $m$-bit symmetric uniform quantization is applied to the network parameters. Quantization-aware training (QAT) is employed during fine-tuning to mitigate quantization errors. Finally, Huffman coding, a lossless entropy coding method, is applied to the quantized parameters to further reduce the overall storage cost. As illustrated in Fig.~\ref{Fig_NeRM3D}, in the compression stage, a neural representation model is overfitted to the 3D RM and then compressed into a compact bitstream for efficient storage and transmission. In the reconstruction stage, the dequantized model weights are loaded into the network for inference to reconstruct the full 3D RM.

\begin{figure}[!t]
\centering
\includegraphics[width=0.39\textwidth]{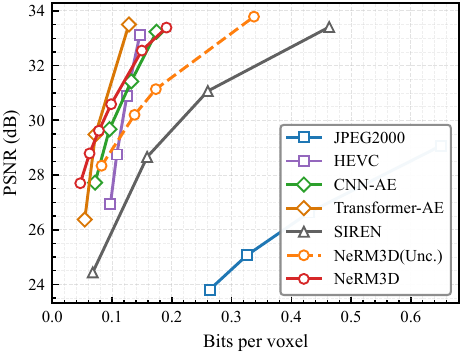}
\caption{The rate-distortion curves of our proposed NeRM3D and different compression baselines in PSNR.}
\label{Fig: rd_performance}
\end{figure}

\section{Evaluation}

This section evaluates the proposed INR-based 3D RM compression framework. The dataset, experimental setup, and evaluation results are presented as follows.
\subsection{Dataset and Evaluation Setup}
We evaluate our method on the UrbanRadio3D dataset~\cite{wang2025radiodiff}. Specifically, we select 50 distinct urban regions from the test set, corresponding to building identifiers 601-650. For each region, one transmitter location is randomly sampled from the 200 available transmitters, resulting in a total of 50 3D RMs for evaluation. 
Following the instance-specific INR paradigm, a separate NeRM3D model is fitted to each 3D RM, with its parameters serving as the compressed representation.
For each selected region, RMs across height and modality dimensions are stacked into a tensor $\mathcal{R} \in \mathbb{R}^{256 \times 256 \times 20 \times 4}$, with a spatial resolution of 1\,m. The channel dimension $C=4$ represents multiple signal modalities, including pathloss, delay, and DoA in both azimuth and elevation~\cite{wang2025radiodiff}.

The NeRM3D model is trained for 50{,}000 iterations, followed by an additional 3{,}000 fine-tuning iterations for QAT. The model is trained with the Adam optimizer with a cosine-annealed learning rate from \(2\times10^{-4}\) to \(1\times10^{-4}\). The quantization bit width is set to $m=10$. For PE, we set \(\beta_z=1.25\), \(\beta_{xy}=1.5\), \(L_z=16\), and \(L_{xy}=40\). The rate--distortion points are obtained by varying the MLP hidden dimension \(d_h\) and initial CondConv channel dimension \(c_0\), with \((d_h,c_0)\in\{(12,8),(16,8),(16,16),(20,16),(24,24),(30,24)\}\). All experiments are conducted on a single NVIDIA RTX 4060 GPU.

We compare NeRM3D with three categories of methods: (i) conventional codecs, including JPEG2000~\cite{skodras2002jpeg} and HEVC~\cite{sullivan2012overview}; (ii) AE-based methods, including convolutional neural network (CNN)- and Transformer-based scale-hyperprior codecs; and (iii) INR-based compression represented by SIREN~\cite{sitzmann2020implicit}. JPEG2000 compresses the 3D RM as a stack of grayscale slices,  whereas HEVC treats height as the temporal dimension. The AE-based methods are trained on 3D RMs from 601 training scenes in the UrbanRadio3D dataset. Performance is evaluated using peak signal-to-noise ratio (PSNR), root mean square error (RMSE), normalized mean square error (NMSE), and bits-per-voxel (BPV).

\subsection{Performance Evaluation}
\begin{figure*}[!t]
\centering
\includegraphics[width=0.85\textwidth]{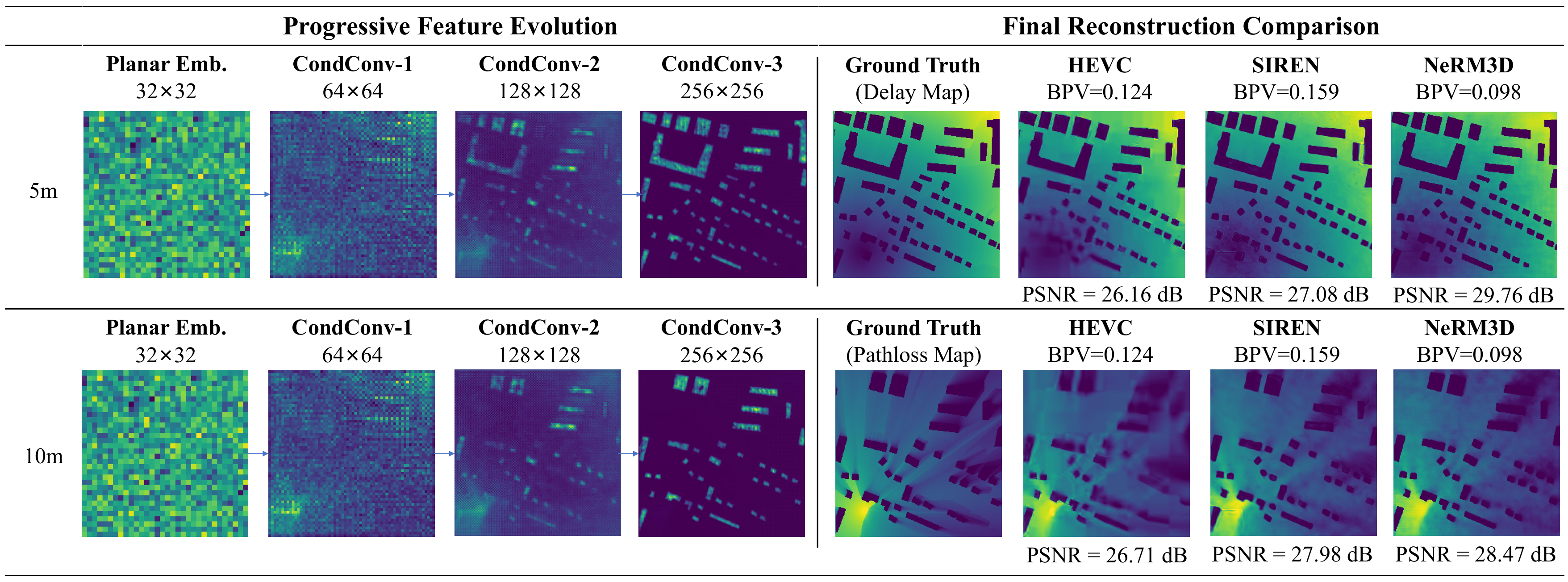}
\caption{Visualization of channel-aggregated feature evolution during progressive upsampling and reconstruction comparisons for radio map slices at different heights.}
\label{Fig: visualize}
\end{figure*}

\begin{table*}[t]
\centering
\caption{Compression performance, decoding latency, and computational complexity under low-bitrate settings}
\label{tab:complexity}
\begin{tabular}{lcccccccc}
\toprule
Method & BPV $\downarrow$ & PSNR $\uparrow$ & NMSE $\downarrow$ & RMSE $\downarrow$
& CPU Latency (ms) $\downarrow$ & GPU Latency (ms) $\downarrow$
& Params $\downarrow$ & MACs $\downarrow$ \\
\midrule
JPEG2000       & 0.263 & 23.81 & 0.0209 & 0.0684 & 289.09  & --     & --       & --        \\
HEVC           & 0.097 & 26.93 & 0.0093 & 0.0464 & 308.28  & --     & --       & --        \\
CNN-AE         & 0.071 & 27.72 & 0.0075 & 0.0420 & 834.76  & 75.52  & 1.3834 M & 33.210 G  \\
Transformer-AE & 0.054 & 26.37 & 0.0096 & 0.0484 & 4826.28 & 409.56 & 2.0319 M & 134.943 G \\
SIREN          & 0.068 & 24.44 & 0.0348 & 0.0730 & 330.21  & 14.36  & 0.0055 M & 7.004 G   \\
\textbf{NeRM3D}& \textbf{0.062} & \textbf{28.79} & \textbf{0.0076}
               & \textbf{0.0397} & \textbf{168.62} & \textbf{13.41}
               & \textbf{0.0090 M} & \textbf{0.783 G} \\
\bottomrule
\end{tabular}
\end{table*}

The rate-distortion performance comparison of different methods is presented in Fig.~\ref{Fig: rd_performance}.  The uncompressed NeRM3D (Unc.) employs the same half-precision (FP16) model representation as the INR-based SIREN baseline, without applying model compression. Conventional image codecs are designed for 2D data and cannot fully exploit cross-height and cross-modal correlations in 3D RMs. NeRM3D consistently outperforms SIREN at comparable bitrates, demonstrating the effectiveness of the proposed height-conditioned representation.
After model compression, compressed NeRM3D further achieves substantially lower bitrates while maintaining competitive reconstruction quality, highlighting the effectiveness of the proposed compression pipeline. In particular, NeRM3D reduces the bitrate from 0.082 to 0.047 BPV with only a minor PSNR degradation from 28.34 dB to 27.70 dB. Moreover, NeRM3D outperforms the mature HEVC codec and the considered AE-based methods in the low-bitrate regime, achieving superior reconstruction quality under stringent compression constraints. Fig.~\ref{Fig: visualize} shows visualizations of reconstructed RM slices for a representative sample. Under a lower memory budget, NeRM3D preserves finer spatial details and achieves better reconstruction quality. These results validate the effectiveness of the proposed INR-based 3D RM compression framework.

Table~\ref{tab:complexity} compares the rate-distortion performance, decoding latency, and computational complexity of different compression methods. Under extremely low bitrate conditions, NeRM3D  achieves superior reconstruction quality over the compared methods. Moreover, NeRM3D achieves approximately $1.8\times$ faster CPU decoding than conventional codecs and $5.6$--$30.5\times$ faster GPU decoding than AE-based methods, while maintaining the lowest latency among the evaluated neural methods. Compared with AE-based methods, INR-based compression enables lightweight decoding without complex two-sided encoder--decoder model collaboration. Furthermore, NeRM3D requires substantially fewer multiply--accumulate operations (MACs) than SIREN by sharing planar representations across heights and avoiding independent voxel-wise coordinate queries. These results demonstrate the computational efficiency of NeRM3D and its suitability for efficient decoding on GPU-equipped devices.

To illustrate the progressive reconstruction process, Fig.~\ref{Fig: visualize} visualizes the intermediate features of NeRM3D across decoding stages. The channel-aggregated feature maps are progressively upsampled through successive CondConv blocks, revealing increasingly refined spatial structures and local variations. The coarse-to-fine design enables high-resolution reconstruction from a compact low-resolution representation while reducing the model size and associated storage overhead compared with direct full-resolution prediction.

\textbf{Remark:} NeRM3D performs offline fitting for each RM during compression. Under the compressed configuration in Table~\ref{tab:complexity}, the encoding time is approximately $74.16$ minutes per RM. For frequently updated RMs, the fitting overhead could be reduced through warm-start or incremental training. For large-scale RM compression, meta-learned initialization, shared-prior INR models, or hypernetwork-based approaches could be explored to reduce the per-instance fitting cost.

\begin{table}[t]
\centering
\caption{Ablation study on key components of NeRM3D}
\label{tab:ablation}
\begin{tabular}{lcc}
\toprule
Configuration & PSNR (dB) & BPV \\
\midrule
NeRM3D (PE + SINE \& InstanceNorm) & 30.59 & 0.098 \\
\midrule
w/o PE & 27.97 & 0.097 \\
PE + GELU & 30.01 & 0.096 \\
\midrule
w/o normalization & 29.64 & 0.098 \\
\midrule
w/o model compression & 31.14 & 0.173 \\
\bottomrule
\end{tabular}
\end{table}

\subsection{Ablation Study}
To evaluate key design choices in NeRM3D, we conduct an ablation study on PE, normalization strategy, and model compression, as summarized in Table~\ref{tab:ablation}.

Removing PE leads to a noticeable performance drop. For activation functions within the PE branch, SINE outperforms GELU, demonstrating stronger capability in modeling high-frequency variations in RMs, particularly around building boundaries. The minor BPV differences stem from entropy coding, as different configurations slightly alter the distribution of quantized model weights.
For normalization in the CondConv block, incorporating instance normalization leads to improved performance compared with the variant without normalization. Finally, removing model compression significantly increases BPV with only marginal PSNR gains.

\section{Conclusion}
In this paper, we propose an implicit neural representation-based framework called NeRM3D to compress 3D RMs. Our key insight is that a 3D RM can be effectively represented by a neural network that takes discrete height values as input and outputs the corresponding RM slices.
Building on this insight, the proposed NeRM3D integrates height modulation with horizontal embeddings, enabling effective extraction and fusion of spatial correlations across different channel modalities in 3D RMs. Based on this representation, we demonstrate that NeRM3D achieves competitive reconstruction performance while enabling fast and lightweight decoding that is well-suited for computation- and memory-constrained UEs. Overall, this work establishes a new paradigm for efficient 3D RM compression in future low-altitude wireless networks based on implicit neural representations.

\ifCLASSOPTIONcaptionsoff
  \newpage
\fi

\bibliographystyle{IEEEtran}
\bibliography{myref}

\end{document}